\documentclass[journal=nalefd,manuscript=letter]{achemso}

\usepackage{amssymb}
\usepackage{amsmath}
\usepackage{chemformula} % Formula subscripts using \ch{}
\usepackage[T1]{fontenc} % Use modern font encodings

\author{Ruslan D. Yamaletdinov}
\affiliation[Novosibirsk State University]
{Novosibirsk State University, Novosibirsk, 630090, Russia}
\alsoaffiliation[Nikolaev Institute]
{Nikolaev Institute of Inorganic Chemistry SB RAS, Novosibirsk, 630090, Russia}
\email{yamaletdinov@niic.nsc.ru}
\author{Yuriy V. Pershin}
\affiliation[usc]
{Department of Physics and Astronomy, University of South Carolina, Columbia, SC 29208, USA}

\title[Ultrafast lithium diffusion in bilayer buckled graphene: A comparative study of Li and Na ]
  {Ultrafast lithium diffusion in bilayer buckled graphene: A comparative study of Li and Na }

\begin{document}

\begin{tocentry}
	\includegraphics{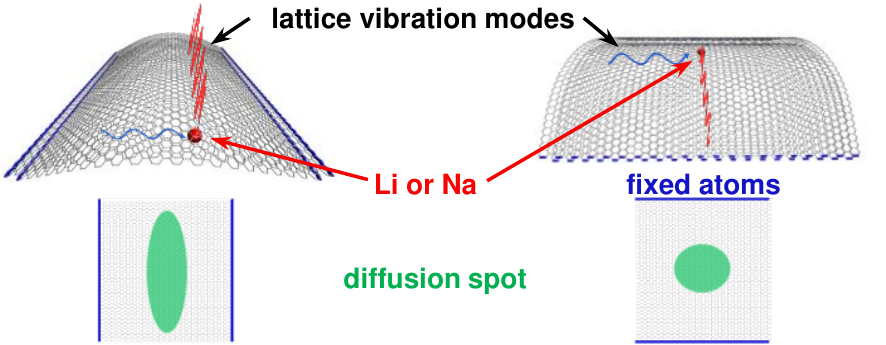}
\end{tocentry}

\begin{abstract}
The effect of the curvature of bilayer  graphene on the interlayer diffusion of Li atoms is investigated using molecular dynamics simulations. A spectacular enhancement of the diffusion constant parallel to the folding axis is found. The ratio of the parallel to the perpendicular diffusion  depends on the buckling direction and stacking type, and it increases with the degree of buckling. The strongest anisotropy  is observed in the case of fixed zig-zag edges.  A comparison with the interlayer diffusion of Na suggests that the strong asymmetry in the  vibrational states of buckled graphene and also the smaller mass of Li are likely to contribute to the observed diffusion enhancement. This work opens a new pathway to develop highly-efficient anodes for rechargeable alkaline batteries.
\end{abstract}
% Our observations can be explained by a strong asymmetry in the vibrational states of buckled graphene.
	%% main text
%	\section{Introduction}
%	\label{intro}
	Lithium-ion batteries~\cite{tarascon2011issues,Nitta2015} have been continuously improved over the past few decades and they have now become the first choice for many consumer electronic devices, zero emission electric vehicles, and even  sustainable energy technology~\cite{Thomas2009,Chu2016}. A further progress in their capacity and cycling performance is directly related to the development of new electrolytes and advanced electrode materials. In particular, it has been shown that various wrapped, coated and crumpled composites~\cite{Choi2014, Chen2017,Nitta2015} offer a superior performance compared to  graphite for application as anode materials.

A fast diffusion of lithium ions through the anode material is one of the requirements to achieve high power, large capacity batteries that are able to operate at fast
charge/discharge rates~\cite{wu2011doped}. In this paper, we demonstrate a spectacular enhancement of the interlayer diffusion of Li in the buckled bilayer graphene. Using molecular dynamics simulations, the anisotropic diffusion constants for the interlayer diffusion are found.
A  physical mechanism is put forward to explain the ultrafast diffusion that is observed in our simulations.
Importantly, as many graphene composites contain curved graphene sheets on their surface, our findings are relevant to real experimental situations. Moreover, the anodes incorporating curved, buckled, or rolled graphene  can easily be fabricated by suitable mechanical means.

	There are many theoretical and experimental studies dedicated to the thermodynamic properties of lithium in graphitic materials. In particular, it was shown that lithium atoms have a fairly high adhesion energy $\sim 1.3$~eV\cite{Fan2012,Koh2013} ($\sim 0.7-1$~eV for Na~\cite{Sun2017,Koh2013}), and that this energy is increasing in the vicinity of defects~\cite{Fan2012,Zhou2012}. This leads to the directional asymmetry of the diffusion barrier.
In pristine graphene, this barrier is $\sim 0.3$~eV for Li~\cite{Uthaisar2010,Zhou2012,Tsai2015a} and  $\sim 0.13$~eV for Na~\cite{Koh2013}. However, when an atom displaces towards a defect (such as a vacancy, divacancy or Stone-Wales defect), the barrier energy may decrease by up to $\sim 60\%$ for Li~\cite{Fan2012} and $\sim 80\%$ for Na~\cite{Yang2016,Yang2017}. A similar effect occurs when an atom moves towards an edge~\cite{Sun2017,Uthaisar2010}.

	Moreover, it has been shown that the diffusion barrier for alkaline atoms absorbed on a monolayer graphene depends on its curvature~\cite{Koh2013}. Density functional theory (DFT) calculations have demonstrated  that the energy barrier decreases/increases if the atom is absorbed from the concave/convex side. In a wide range, for any type of buckling, the energy barrier for Na atoms is about $0.1$~eV lower than that for Li atoms~\cite{Koh2013}.
Ref. \citenum{Koh2013} also reports a diffusion anisotropy with a variation of the energy barrier of $0.08$~eV for Li and $0.03$~eV for Na.
 %The authors do not provide any additional analysis and do not try to separate the influence of geometric and electronic interactions on the resulting barrier energies.

However, it is more difficult to make a similar analysis for the diffusion in the bilayer graphene because of a larger number of degrees of freedom.
 The effect of the interlayer distance on Li and Na intercalation and diffusion was evaluated
in Refs. \citenum{Xu2012} and \citenum{Tsai2015a}, respectively. Recently, Zhong et. al.~\cite{Zhong2019} performed a series of \textit{ab initio} calculations of lithium diffusion in  AA and AB stacked graphene. According to these calculations, in the case of AA stacking, the diffusion is isotropic, and the energy barrier is $0.34$~eV. In the case of AB stacking, there are two distinct possibilities for lithium migration with the barrier heights of $0.07$  and $0.25$~eV. However, these previous results neglect the impact of the vibrational states on diffusion, which may be of significant importance~\cite{Lebedeva2011}.

%	\section{Calculation details}
%	\label{sim_det}
	Molecular dynamics (MD) simulations were carried out with NAMD2 software package~\cite{phillips05} (NAMD was developed by the Theoretical and Computational Biophysics Group in the Beckman Institute for Advanced Science and Technology at the University of Illinois at Urbana-Champaign).
	To describe the interaction between carbon atoms, the CHARMM-like potential (including standard 2-body spring bond,
	3-body angular bond, including the Urey-Bradley term, 4-body torsion angle and Lennard-Jones potential energy terms) that was previously optimized for graphene~\cite{Yamaletdinov18} was used. Knowing that the lithium-carbon interaction can be quite accurately described by the linear superposition of power-law functions of the type $1/r^{n}$~\cite{Khantha2004}, the Lennard-Jones potential was employed to approximate Li-C and Na-C interactions. The Li-C bonding parameters were optimized to fit the equilibrium distance and energy from \textit{ab-initio} calculations~\cite{Khantha2004}. A similar procedure was carried out to fit the \textit{ab-initio} Na-C distance~\cite{Tsai2015b} and energy~\cite{Koh2013}. The Van-der-Waals interactions were gradually cut off, starting at $10$\r{A} from the atom until reaching zero $12$\r{A} away.
	 	
	MD simulations were performed with 1~fs time step. The Langevin dynamics with a damping parameter of $5 \textnormal{ps}^{-1}$ was used for the temperature control. In all our simulations, the energy was first minimized in 2000 steps. Next, the system dynamics was simulated for $\tau=0.5$ns. Finally, the energy was minimized in 2000 steps.

	\begin{figure}[tb]
		\includegraphics[width=75mm]{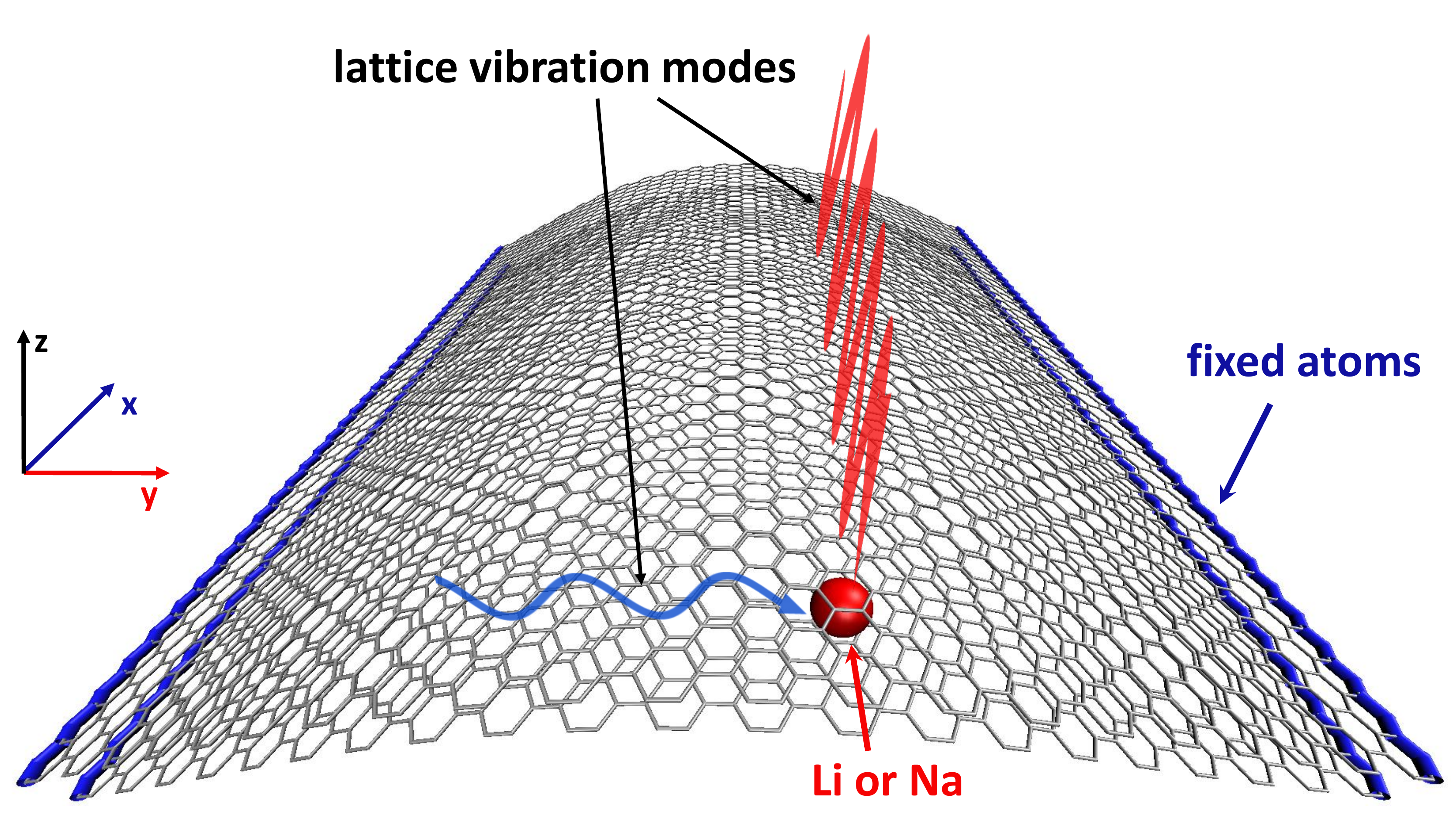}
		\caption{Numerical simulation setup.}
        \label{fig:1}
	\end{figure}

	Fig.~\ref{fig:1} shows a typical simulation setup. Here, an initially optimized bilayer graphene sheet of $L=93.62$\r{A} length (along $x$-axis) and $W=86.82$~\r{A} width (along $y$-axis) is buckled around $x$-axis.  In each simulation, a Li or Na atom was placed between the layers of AA- or AB-stacked bilayer graphene in a random initial position.
	The buckling was carried out by fixing the edge atoms at a distance $d<L$ (in case of fixing armchair edge (AC)) or $d<W$ (in case of fixing zig-zag edge (ZZ)). The role of the temperature was analyzed by performing simulations at three different temperatures (with a step of $50$K) for each graphene geometry. At each temperature, the statistics were collected in 250 runs. To correlate our results with data available in the literature, the alkaline diffusion coefficients were also found for a monolayer (1L) and bilayer (2L) graphene without any fixed atoms.

Anisotropic diffusion coefficients at the temperature $T$, $D_T^\parallel$ and $D_T^\perp$, were determined using the mean square displacement (MSD). Here, $\parallel$ is the direction about which the buckling takes place ($x$ in Fig. \ref{fig:1}) and $\perp$ is the complementary direction  ($y$ in Fig. \ref{fig:1}). In the case of Fig. \ref{fig:1},
	\begin{equation}
	\label{eq:d_msd}
		D_T^\parallel  = \sum_{i=1}^N \frac{(x_i(\tau)-x_i(0))^2}{2 N\tau},\qquad
		D_T^\perp  = \sum_{i=1}^N \frac{(y_i(\tau)-y_i(0))^2}{2 N\tau},
	\end{equation}
%	\begin{equation*}
%		D=\frac{1}{2}(D_x+D_y),
%	\end{equation*}
	where $x_i(\tau)$ and $y_i(\tau)$ are the coordinates of Li or Na atom at the final moment of time $\tau$ in the $i$-th run, $x_i(0)$ and $y_i(0)$ are the initial coordinates, and $N=250$ is the  number of runs.
	
The temperature dependence of diffusion was fitted by the Arrhenius equation
	\begin{equation}
	\label{eq:d_exp}
		D_T^{\parallel(\perp)} = Z_{\parallel (\perp)} \exp \left(-\frac{E_{\parallel(\perp)}}{k T}\right),
	\end{equation}
	where $Z_{\parallel(\perp)}$ is the frequency factor~\cite{dienes1950frequency} in the $\parallel$ (or $\perp$) direction,
	$E_{\parallel(\perp)}$ is the corresponding activation energy, and $k$ is the Boltzmann constant. In all simulated cases, MD results were fitted by Eq.~\ref{eq:d_exp}  with a RMS error less then $1\%$.

		 \begin{table}[b]
	 	\begin{center}
	 		\caption{Li diffusion coefficient at $T=300$~K, frequency factor, and activation energy in single-layer (1L) and AA-stacked bilayer (2L) flat graphene.}
	 		\label{tab:diff_isotropLi}
	 		\begin{tabular}{|c||c|c|c||c|}
	 			\hline
	 			system&
	 			$D_{300}$, cm$^2$/s& $Z$, cm$^2$/s&$E$, eV& prior work \\
	 			\hline
	 			
	 			Li-1L& $6.6\cdot10^{-7}$& $4.7\cdot10^{-3}$&$0.23$&$E=0.28-0.33$~eV ~\cite{Uthaisar2010,Zhou2012,Tsai2015a}  \\
	 			\hline
	 			AA-Li-2L& $6.0\cdot10^{-6}$& $1.6\cdot10^{-4}$ &$0.085$ &$D=4.4 \cdot10^{-6} - 7\cdot 10^{-5}$~cm$^2$/s~\cite{Persson2010,Kuhne2017}\\

	 			\hline
	 		\end{tabular}
	 	\end{center}
	 \end{table}

To correlate our results with data available in the literature, the diffusion in flat graphene was simulated. The results of these calculations for Li are presented in the table \ref{tab:diff_isotropLi}, while for Na - in the Supporting Information (SI) table~\ref{tab:diff_isotropNa} and Fig. S.1.
Our general observations are that the sodium diffusion occurs more easily than  lithium diffusion, and that the interlayer diffusion is more efficient than the surface diffusion. Fig. S.1
demonstrate an isotropic diffusion with MSD increasing with the temperature, which was exactly the expected behavior.
Overall, the numerical diffusion parameters listed in  tables~\ref{tab:diff_isotropLi} and S.1
are in good agreement with the experimental and numerical literature data.
An order of magnitude deviation from the ultrafast experimental value $D_{uf}=7\cdot 10^{-5}$~cm$^2$/s could be related to various factors in the experimental setup~\cite{Kuhne2017} not taken into account in our simple simulation. The authors of Ref.~\citenum{Kuhne2017} also report lower values for $D$ (e.g., in the first lithiation cycle $D=5\cdot 10^{-6}$~cm$^2$/s), and associate its further growth with the establishment of a Li intercalation pathway.
Meanwhile, because the diffusion constant in buckled graphene can be of the same order in magnitude as $D_{uf}$, an (effective) buckling is a possible explanation for the observed ultrafast diffusion constant~\cite{Kuhne2017}.

%	\section{Results}
%	\label{res}

	\begin{figure}[h]
		\includegraphics[width=.9\textwidth]{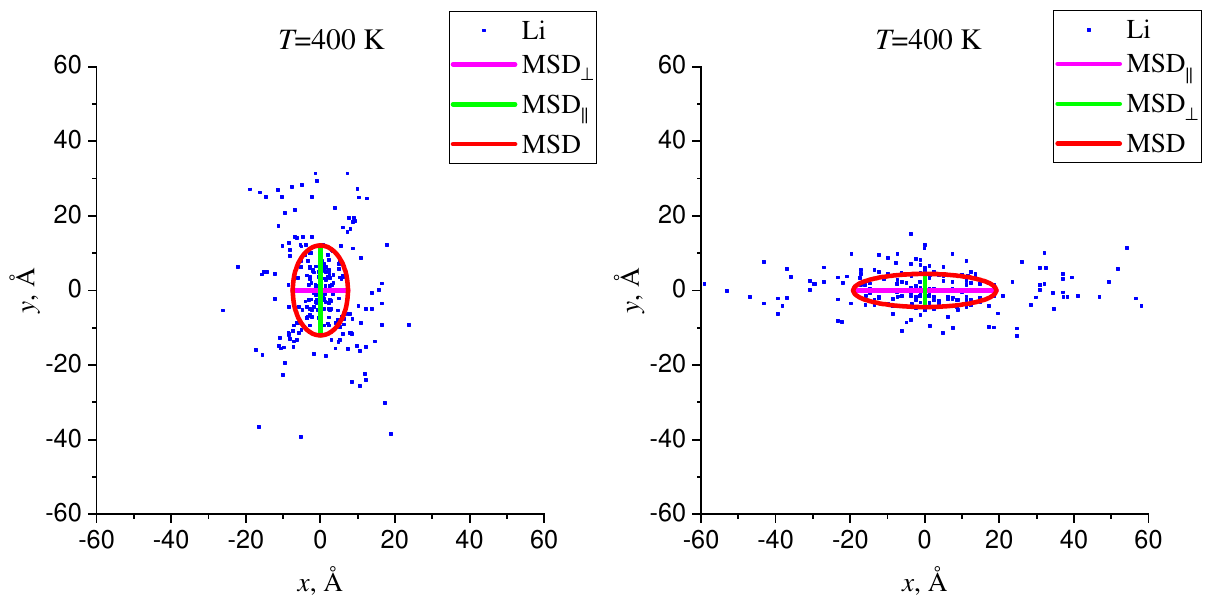}

		\caption{The final displacement of Li atoms in AA-stacked bilayer graphene. The $\parallel$ direction is along the armchair edge (left-hand plot), and zig-zag edge (right-hand plot). The green and purple lines corresponds to the root MSD in $x$ and $y$ directions, respectively.}
		\label{Li_disp}
	\end{figure}
	
	MD simulations for the diffusion in buckled structures were performed using the same procedure as for the flat ones. Some general observations are that the diffusion barrier for Na is slightly lower than that for lithium, the preferable diffusion direction is along the fixed edge (the $\parallel$ direction), and the greater the curvature of graphene the more pronounced is the diffusion anisotropy. For both Li (Fig. \ref{Li_disp})
and Na (SI Fig. S.2%\ref{fig:S2}
), the anisotropy of diffusion in AA-stacked graphene is stronger for the case of fixed zig-zag edges. For Li, a similar result was found for the case of AB stacking (see the last two lines in the table~\ref{tab:diff_Li}).

Our most striking finding is the spectacular increase in the $\parallel$ diffusion constant of Li atoms in buckled structures. Here, we refer to the results for AA-Li-ZZ-90 and AB-Li-ZZ-95 structures (table~\ref{tab:diff_Li}) showing about 30-fold increase in the diffusion compared to the flat single-layer graphene or 3-fold increase compared to the flat bilayer graphene (table~\ref{tab:diff_isotropLi}). Contrastingly, a high diffusivity of Na in the flat bilayer graphene (SI table S.1%\ref{tab:diff_isotropNa}
) decreases with buckling, and is close to the ``flat'' value only in the AA-Na-ZZ-90 system (SI table S.2).
% \ref{tab:diff_Na}).

Naively, one can expect an increase in the frequency factor ($Z{_{\parallel (\perp)}}$) and  diffusion energy barrier ($E_{{\parallel (\perp)}}$) with buckling (due to the higher rigidity of buckled structures). In reality, however, this expectation was not met. The numerical constants presented in tables \ref{tab:diff_Li} and S.2 %\ref{tab:diff_Na}
show a mixed behavior. Finally, the energy barrier in the flat undistorted bilayer graphene (with all fixed carbon atoms) was identified by scanning the coordinate between two nearest local minima. It was found that $E=2.5$~eV for Li and $E=16$~eV for Na at the interlayer distance of $h=3.4$~\r{A}. However, this barrier does not correspond to any real processes because in this calculation all of the carbon atoms were kept fixed.

 %we expected that the more curvature the greater the system rigidity and as a result the greater. In fact, our expectations were not met (see SI.Table~\ref{tab:diff_Na}). In case of fixed zig-zag edge (which is parallel to the $X$ axis), energy barrier along $X$ axis ($E_x$) decreases with increasing curvature. In case of fixed armchair edge Na diffusion parameters met our expectations, unlike Li whose diffusion parameters behave completely contrary to expectations.

		\begin{table}[h!]
		\begin{center}
			\caption{Li diffusion constants at $T=300$~K, frequency factors, and activation energies in  AA- and AB-stacked bilayer graphene. Here AC (ZZ) denotes the type of the edge that was kept fixed in simulations (the edge in the $\parallel$ direction). The number 90 (or 95) in the system name denotes the degree of buckling (i.e., 95 corresponds to $d/L=0.95$).}
			\label{tab:diff_Li}
			\begin{tabular}{|c||c|c|c||c|c|c|}
				\hline
				system&
				$D_{300}^\parallel$, cm$^2$/s&$Z_\parallel$, cm$^2$/s & $E_\parallel$, eV& $D_{300}^\perp$, cm$^2$/s&$Z_\perp$, cm$^2$/s& $E_\perp$, eV\\
				\hline				
				AA-Li-AC-95& $3.2 \cdot 10^{-6}$ &$2.2\cdot 10^{-4}$ &$0.11$ & $2.7 \cdot 10^{-6}$& $1.5\cdot 10^{-4}$ &$0.10$  \\
				\hline
				AA-Li-AC-90& $6.4 \cdot 10^{-6}$ &$1.4\cdot 10^{-4}$ &$0.079$ & $2.1 \cdot 10^{-6}$& $0.98\cdot 10^{-4}$ &$0.10$ \\
				\hline
				AA-Li-ZZ-95& $2.3 \cdot 10^{-6}$& $1.0\cdot 10^{-4}$ &$0.099$ &$3.4 \cdot 10^{-6}$ &$0.50\cdot 10^{-4}$ &$0.070$ \\
				\hline
				AA-Li-ZZ-90& $2.0 \cdot 10^{-5}$ &$2.8\cdot 10^{-4}$ &$0.068$&$5.7 \cdot 10^{-7}$ &$0.83\cdot 10^{-4}$ &$0.13$ \\
                \hline
				AB-Li-AC-95& $5.8 \cdot 10^{-6}$ &$2.4\cdot 10^{-4}$ &$0.96$& $2.7 \cdot 10^{-6}$& $0.99\cdot 10^{-4}$ &$0.093$ \\
                \hline
				AB-Li-ZZ-95& $1.7 \cdot 10^{-5}$& $2.0\cdot 10^{-4}$ &$0.063$&$6.1 \cdot 10^{-7}$ &$0.48\cdot 10^{-4}$ &$0.11$\\
				\hline
			\end{tabular}
		\end{center}
	\end{table}

%	\section{Discussion}
%	\label{disc}
	
\begin{figure}[tb]
		\includegraphics[width=0.9\textwidth]{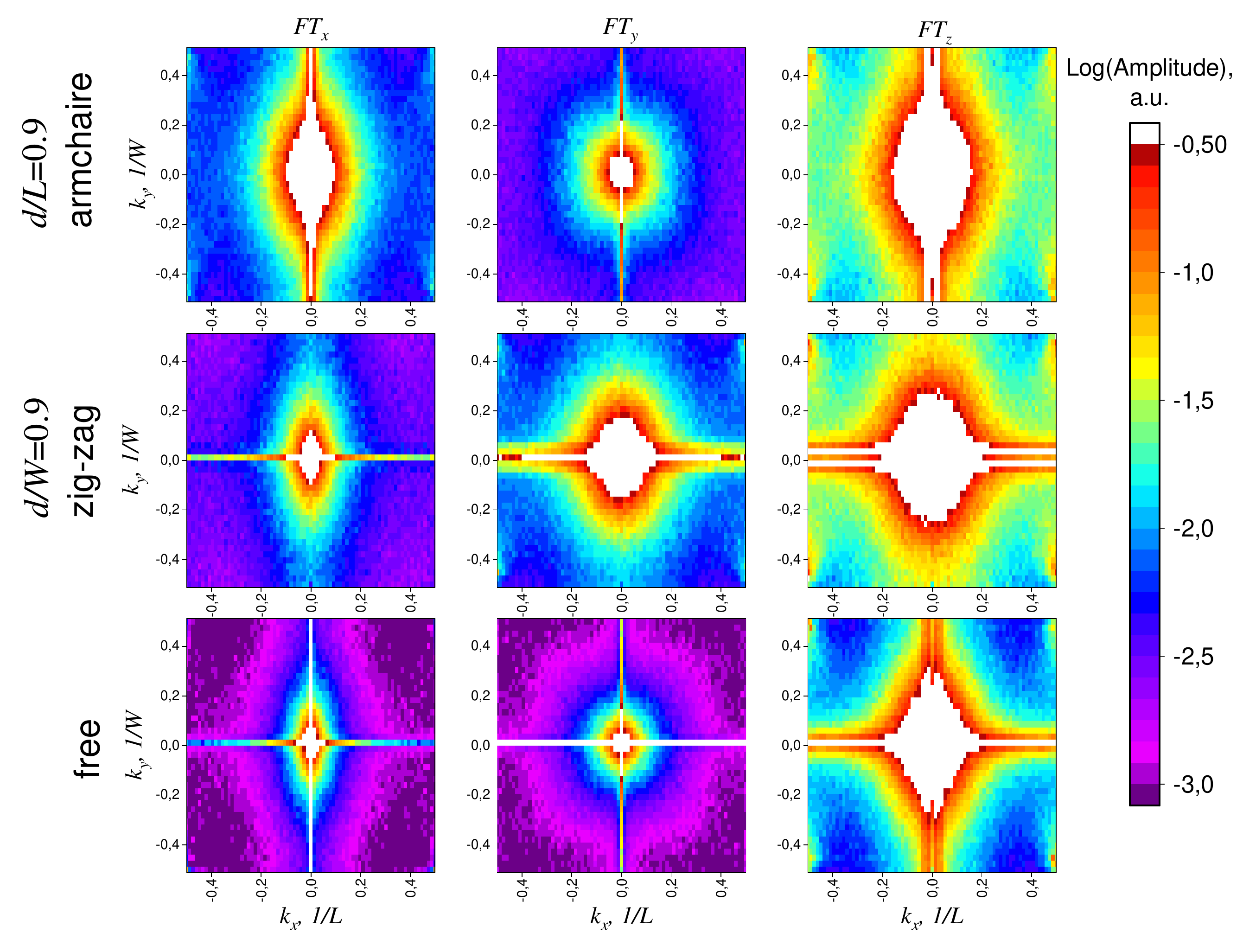}
		\caption{Time-averaged Fourier transform of carbon displacements (Eq.~(\ref{eq:FT_Q})). From left to right: the Fourier transform along $x$, $y$ and $z$ directions. Top row: AA-stacked buckled bilayer graphene with fixed AC edges that are parallel to $y$-axis. Middle row: AA-stacked buckled bilayer graphene with fixed ZZ edges  that are  parallel to $x$-axis. Bottom row: AA-stacked free bilayer graphene. The colour scale is logarithmic. These plots are obtained using parameter values $M=77$, $N=42$, $\Delta t=10$~fs, $J \Delta t\equiv T_0=10$~ps.}
		\label{ft_disp}
	\end{figure}

	It is well-known that the mechanical properties of graphene are anisotropic ~\cite{Ni2010}. The buckling by itself is another source of anisotropy. To reveal the asymmetry of lattice vibrations in the systems under consideration,  Fig.~\ref{ft_disp} exhibits the time-averaged Fourier transform of carbon displacements
	\begin{equation}
	\label{eq:FT_Q}
	%FT_x (k_x ,k_y) = \frac{1}{J} \sum_{j=0}^{J-1} \text{abs}\left[\sum_{n=0}^{N-1} \sum_{m=0}^{M-1} x_{nm}(j \cdot \Delta t) \cdot \exp \left(2\pi i \left[\frac{k_x}{M}m+\frac{k_y}{N}n\right]\right)\right],
FT_x (k_x ,k_y) = \frac{1}{J} \sum_{j=0}^{J-1}\text{abs}\left[\sum_{n=0}^{N-1} \sum_{m=0}^{M-1} x_{mn}(j \cdot \Delta t) e^{2\pi i\left[ k_xmL+k_ynW\right]} \right] ,
	\end{equation}
	where $J$ is the number of time steps, $M$ and $N$  define the grid size (each grid point corresponds to an atom), $x_{mn}(j \cdot \Delta t)$ is the atomic $x$-coordinate  at the grid point $(m,n)$ at time $j \cdot \Delta t$, $\Delta t$ is the time step, $k_x=\tilde{m}/(LM)$, $k_y=\tilde{n}/(WN)$), $\tilde{m}$ and $\tilde{n}$ are integers, and $\text{abs}[...]$ denotes the absolute value. $FT_y$ and $FT_z$ were plotted using similar expressions.

Fig.~\ref{ft_disp} reveals high intensity swing modes at $\vec{k}\approx(0,0)$ in each graph. In the case of the free bilayer graphene, two symmetric lines of high amplitude oscillations (defined by $k_x=0$ and $k_y=0$) are also visible. In buckled structures, however, the vibrations in the $\perp$ direction are suppressed. In particular, Fig.~\ref{ft_disp} shows high-intensity vibrations of only about $k_x\approx0$ when $x-$edges are in the $\parallel$ direction (AC case), and about $k_y\approx0$ when the $y-$edges are in the $\parallel$ direction (ZZ case). This implies
 a strong correlation between the symmetry of high intensity vibrations and diffusion anisotropy.
Given that the energy barrier calculated at all carbon atoms fixed is much higher than that from MD simulations at finite temperatures ($2.5$~eV versus 0.085~eV for Li at 300~K in free graphene),
the physical displacement of carbon atoms is of crucial importance for the ultrafast diffusion discovered in our MD simulations.

In the buckled structures, while $k_\perp$ modes are suppressed, the oscillations with high-$k$ values ($|k_x|\approx0.5$, $|k_y|\approx0.4$) are enhanced.  The wavelengths of these high-$k$ values modes are comparable to few lattice periods. Because such vibrations do strongly modify the order in the $x$ direction, and this effect is most noticeable for the case of fixed ZZ edges, one can associate the decrease in the energy barrier as a function of the curvature in ZZ systems with enhancement of such high-$k_\parallel$ vibrations.

	Being lighter than Na, Li atoms must be affected more by lattice vibrations. To verify this statement, we calculate the frequency dependence of the oscillation density
	\begin{equation}
		\label{eq:FT_t}
	 FT_x(\nu)=\frac{1}{J}\text{abs}\left[ \sum_{j=0}^{J-1}x(j \cdot \Delta t)\cdot \exp (i2\pi\cdot\nu \cdot j \Delta t )\right],
%FT_x(\nu_n)=\frac{1}{T_0}\text{abs}\int\limits_0^{T_0}x(t)e^{2\pi i\nu_n t}\textnormal{d}t,
	\end{equation}
	where $\nu_n=n/T_0$ is the frequency ($n$ is an integer), $x(j \cdot \Delta t)$ is the $x$-coordinate of an atom at time $j \cdot \Delta t$.  In our work, the time-averaging was performed with the time step $\Delta t$. $FT_y$ and $FT_z$ were found using similar expressions.
 We note that the oscillation frequencies calculated based on Eq. (\ref{eq:FT_t}) are in a good agreement with known DFT results~\cite{Kaneko2017}.

\begin{figure}[h]
		\includegraphics[width=.6\textwidth]{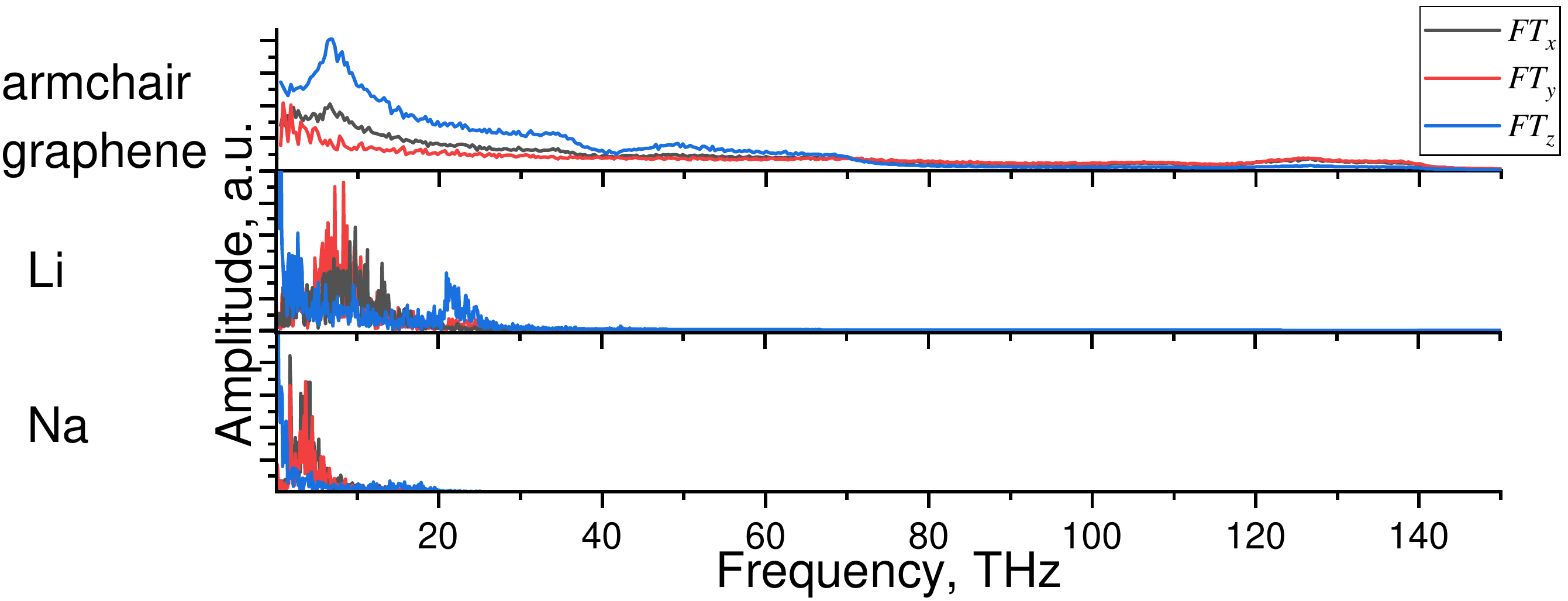}
		\caption{Oscillation density (Eq.~(\ref{eq:FT_t})) for the carbon (top), lithium (middle) and sodium (bottom) atoms. The alkali atoms frequencies were calculated using a free bilayer graphene. }
		\label{ft_at}
	\end{figure}

Fig.~\ref{ft_at} shows  that the frequencies of lithium are higher than these of sodium. Interestingly,  the most intense Li peak matches the peak of graphene. This frequency match could be associated either with the small mass of lithium or with a parametric resonance through which the lattice vibrations pump the lithium atoms.  For some (not completely clear) reasons, the highest amplitude of Li oscillations is observed in $y$-direction, while the peak of oscillations along $x$-axis has a higher frequency. Whatever the reason (the resonance or small mass), the same prevailing frequencies of lithium and graphene oscillations provide in-phase oscillation condition that might contribute to the diffusion barrier decrease.
	
%	\section{Final Remarks}
%	\label{fin}

 Our MD simulations have demonstrated a strong anisotropy of Li diffusion in the buckled bilayer graphene. Our main finding is
 the ultrafast diffusion of Li atoms in buckled structures with fixed zig-zag edges. In particular, the molecular dynamics simulations presented in this work have shown about 30-fold increase
 in the Li diffusion in buckled graphene compared  to single-layer graphene (or about 3-fold increase compared to the free bilayer graphene).
 This effect increases with the degree of buckling and seems to be independent on the stacking type (although only AA- and AB-stacking were considered).
 Qualitatively, our findings could be explained by peculiarities of lattice vibrations in buckled structures possibly assisted by a parametric resonance.

Our work may facilitate the development of more efficient Li-ion batteries, and could be useful in various other applications and studies related to the diffusion of alkaline atoms.

	\begin{acknowledgement}
	
	This work has been partially supported by the Russian Science Foundation grant No. 19-73-10068.
	\end{acknowledgement}
	
%	\begin{suppinfo}
%		The supporting information file contains figures and tables such as the final displacement of Na atoms in single layer graphene (Fig. S.1) and AA-stacked bilayer graphene (Fig. S.2), and Na diffusion parameters for flat geometries (table S.1) and  AA-stacked buckled bilayer graphene (table S.2).
%	\end{suppinfo}
	\bibliography{biblio}

\providecommand{\latin}[1]{#1}
\providecommand*\mcitethebibliography{\thebibliography}
\csname @ifundefined\endcsname{endmcitethebibliography}
  {\let\endmcitethebibliography\endthebibliography}{}
\begin{mcitethebibliography}{28}
\providecommand*\natexlab[1]{#1}
\providecommand*\mciteSetBstSublistMode[1]{}
\providecommand*\mciteSetBstMaxWidthForm[2]{}
\providecommand*\mciteBstWouldAddEndPuncttrue
  {\def\EndOfBibitem{\unskip.}}
\providecommand*\mciteBstWouldAddEndPunctfalse
  {\let\EndOfBibitem\relax}
\providecommand*\mciteSetBstMidEndSepPunct[3]{}
\providecommand*\mciteSetBstSublistLabelBeginEnd[3]{}
\providecommand*\EndOfBibitem{}
\mciteSetBstSublistMode{f}
\mciteSetBstMaxWidthForm{subitem}{(\alph{mcitesubitemcount})}
\mciteSetBstSublistLabelBeginEnd
  {\mcitemaxwidthsubitemform\space}
  {\relax}
  {\relax}

\bibitem[Tarascon and Armand(2011)Tarascon, and Armand]{tarascon2011issues}
Tarascon,~J.-M.; Armand,~M. \emph{Materials for Sustainable Energy: A
  Collection of Peer-Reviewed Research and Review Articles from Nature
  Publishing Group}; World Scientific, 2011; pp 171--179\relax
\mciteBstWouldAddEndPuncttrue
\mciteSetBstMidEndSepPunct{\mcitedefaultmidpunct}
{\mcitedefaultendpunct}{\mcitedefaultseppunct}\relax
\EndOfBibitem
\bibitem[Nitta \latin{et~al.}(2015)Nitta, Wu, Lee, and Yushin]{Nitta2015}
Nitta,~N.; Wu,~F.; Lee,~J.~T.; Yushin,~G. \emph{Materials Today} \textbf{2015},
  \emph{18}, 252--264\relax
\mciteBstWouldAddEndPuncttrue
\mciteSetBstMidEndSepPunct{\mcitedefaultmidpunct}
{\mcitedefaultendpunct}{\mcitedefaultseppunct}\relax
\EndOfBibitem
\bibitem[Thomas(2009)]{Thomas2009}
Thomas,~C. \emph{International Journal of Hydrogen Energy} \textbf{2009},
  \emph{34}, 6005--6020\relax
\mciteBstWouldAddEndPuncttrue
\mciteSetBstMidEndSepPunct{\mcitedefaultmidpunct}
{\mcitedefaultendpunct}{\mcitedefaultseppunct}\relax
\EndOfBibitem
\bibitem[Chu \latin{et~al.}(2016)Chu, Cui, and Liu]{Chu2016}
Chu,~S.; Cui,~Y.; Liu,~N. \emph{Nature Materials} \textbf{2016}, \emph{16},
  16--22\relax
\mciteBstWouldAddEndPuncttrue
\mciteSetBstMidEndSepPunct{\mcitedefaultmidpunct}
{\mcitedefaultendpunct}{\mcitedefaultseppunct}\relax
\EndOfBibitem
\bibitem[Choi and Kang(2014)Choi, and Kang]{Choi2014}
Choi,~S.~H.; Kang,~Y.~C. \emph{ChemSusChem} \textbf{2014}, \emph{7},
  523--528\relax
\mciteBstWouldAddEndPuncttrue
\mciteSetBstMidEndSepPunct{\mcitedefaultmidpunct}
{\mcitedefaultendpunct}{\mcitedefaultseppunct}\relax
\EndOfBibitem
\bibitem[Chen \latin{et~al.}(2017)Chen, Shen, van Aken, Maier, and
  Yu]{Chen2017}
Chen,~S.; Shen,~L.; van Aken,~P.~A.; Maier,~J.; Yu,~Y. \emph{Advanced
  Materials} \textbf{2017}, \emph{29}, 1605650\relax
\mciteBstWouldAddEndPuncttrue
\mciteSetBstMidEndSepPunct{\mcitedefaultmidpunct}
{\mcitedefaultendpunct}{\mcitedefaultseppunct}\relax
\EndOfBibitem
\bibitem[Wu \latin{et~al.}(2011)Wu, Ren, Xu, Li, and Cheng]{wu2011doped}
Wu,~Z.-S.; Ren,~W.; Xu,~L.; Li,~F.; Cheng,~H.-M. \emph{ACS nano} \textbf{2011},
  \emph{5}, 5463--5471\relax
\mciteBstWouldAddEndPuncttrue
\mciteSetBstMidEndSepPunct{\mcitedefaultmidpunct}
{\mcitedefaultendpunct}{\mcitedefaultseppunct}\relax
\EndOfBibitem
\bibitem[Fan \latin{et~al.}(2012)Fan, Zheng, and Kuo]{Fan2012}
Fan,~X.; Zheng,~W.; Kuo,~J.-L. \emph{ACS Applied Materials {\&} Interfaces}
  \textbf{2012}, \emph{4}, 2432--2438\relax
\mciteBstWouldAddEndPuncttrue
\mciteSetBstMidEndSepPunct{\mcitedefaultmidpunct}
{\mcitedefaultendpunct}{\mcitedefaultseppunct}\relax
\EndOfBibitem
\bibitem[Koh and Manzhos(2013)Koh, and Manzhos]{Koh2013}
Koh,~Y.~W.; Manzhos,~S. \emph{MRS Communications} \textbf{2013}, \emph{3},
  171--175\relax
\mciteBstWouldAddEndPuncttrue
\mciteSetBstMidEndSepPunct{\mcitedefaultmidpunct}
{\mcitedefaultendpunct}{\mcitedefaultseppunct}\relax
\EndOfBibitem
\bibitem[Sun \latin{et~al.}(2017)Sun, Wang, and Fu]{Sun2017}
Sun,~X.; Wang,~Z.; Fu,~Y.~Q. \emph{Carbon} \textbf{2017}, \emph{116},
  415--421\relax
\mciteBstWouldAddEndPuncttrue
\mciteSetBstMidEndSepPunct{\mcitedefaultmidpunct}
{\mcitedefaultendpunct}{\mcitedefaultseppunct}\relax
\EndOfBibitem
\bibitem[Zhou \latin{et~al.}(2012)Zhou, Hou, and Wu]{Zhou2012}
Zhou,~L.~J.; Hou,~Z.~F.; Wu,~L.~M. \emph{Journal of Physical Chemistry C}
  \textbf{2012}, \emph{116}, 21780--21787\relax
\mciteBstWouldAddEndPuncttrue
\mciteSetBstMidEndSepPunct{\mcitedefaultmidpunct}
{\mcitedefaultendpunct}{\mcitedefaultseppunct}\relax
\EndOfBibitem
\bibitem[Uthaisar and Barone(2010)Uthaisar, and Barone]{Uthaisar2010}
Uthaisar,~C.; Barone,~V. \emph{Nano Letters} \textbf{2010}, \emph{10},
  2838--2842\relax
\mciteBstWouldAddEndPuncttrue
\mciteSetBstMidEndSepPunct{\mcitedefaultmidpunct}
{\mcitedefaultendpunct}{\mcitedefaultseppunct}\relax
\EndOfBibitem
\bibitem[Tsai \latin{et~al.}(2015)Tsai, Chung, Lin, and Yamada]{Tsai2015a}
Tsai,~P.-c.; Chung,~S.-C.; Lin,~S.-k.; Yamada,~A. \emph{Journal of Materials
  Chemistry A} \textbf{2015}, \emph{3}, 9763--9768\relax
\mciteBstWouldAddEndPuncttrue
\mciteSetBstMidEndSepPunct{\mcitedefaultmidpunct}
{\mcitedefaultendpunct}{\mcitedefaultseppunct}\relax
\EndOfBibitem
\bibitem[Yang \latin{et~al.}(2016)Yang, Li, Tang, Dong, Sun, Shen, and
  Wang]{Yang2016}
Yang,~S.; Li,~S.; Tang,~S.; Dong,~W.; Sun,~W.; Shen,~D.; Wang,~M.
  \emph{Theoretical Chemistry Accounts} \textbf{2016}, \emph{135}, 164\relax
\mciteBstWouldAddEndPuncttrue
\mciteSetBstMidEndSepPunct{\mcitedefaultmidpunct}
{\mcitedefaultendpunct}{\mcitedefaultseppunct}\relax
\EndOfBibitem
\bibitem[Yang \latin{et~al.}(2017)Yang, Li, Tang, Shen, Dong, and
  Sun]{Yang2017}
Yang,~S.; Li,~S.; Tang,~S.; Shen,~D.; Dong,~W.; Sun,~W. \emph{Surface Science}
  \textbf{2017}, \emph{658}, 31--37\relax
\mciteBstWouldAddEndPuncttrue
\mciteSetBstMidEndSepPunct{\mcitedefaultmidpunct}
{\mcitedefaultendpunct}{\mcitedefaultseppunct}\relax
\EndOfBibitem
\bibitem[Xu \latin{et~al.}(2012)Xu, Wu, Liu, and Ouyang]{Xu2012}
Xu,~B.; Wu,~M.~S.; Liu,~G.; Ouyang,~C.~Y. \emph{Journal of Applied Physics}
  \textbf{2012}, \emph{111}\relax
\mciteBstWouldAddEndPuncttrue
\mciteSetBstMidEndSepPunct{\mcitedefaultmidpunct}
{\mcitedefaultendpunct}{\mcitedefaultseppunct}\relax
\EndOfBibitem
\bibitem[Zhong \latin{et~al.}(2019)Zhong, Hu, Xu, Yang, Zhang, and
  Huang]{Zhong2019}
Zhong,~K.; Hu,~R.; Xu,~G.; Yang,~Y.; Zhang,~J.-M.; Huang,~Z. \emph{Physical
  Review B} \textbf{2019}, \emph{99}, 155403\relax
\mciteBstWouldAddEndPuncttrue
\mciteSetBstMidEndSepPunct{\mcitedefaultmidpunct}
{\mcitedefaultendpunct}{\mcitedefaultseppunct}\relax
\EndOfBibitem
\bibitem[Lebedeva \latin{et~al.}(2011)Lebedeva, Knizhnik, Popov, Lozovik, and
  Potapkin]{Lebedeva2011}
Lebedeva,~I.~V.; Knizhnik,~A.~A.; Popov,~A.~M.; Lozovik,~Y.~E.; Potapkin,~B.~V.
  \emph{Physical chemistry chemical physics : PCCP} \textbf{2011}, \emph{13},
  5687--5695\relax
\mciteBstWouldAddEndPuncttrue
\mciteSetBstMidEndSepPunct{\mcitedefaultmidpunct}
{\mcitedefaultendpunct}{\mcitedefaultseppunct}\relax
\EndOfBibitem
\bibitem[Phillips \latin{et~al.}(2005)Phillips, Braun, Wand, Gumbart,
  Tajkhorshid, Villa, Chipot, Skeel, Kale, and Schulten]{phillips05}
Phillips,~J.~C.; Braun,~R.; Wand,~W.; Gumbart,~J.; Tajkhorshid,~E.; Villa,~E.;
  Chipot,~C.; Skeel,~R.~D.; Kale,~L.; Schulten,~K. \emph{J. Comp. Chem.}
  \textbf{2005}, \emph{26}, 1781--1802\relax
\mciteBstWouldAddEndPuncttrue
\mciteSetBstMidEndSepPunct{\mcitedefaultmidpunct}
{\mcitedefaultendpunct}{\mcitedefaultseppunct}\relax
\EndOfBibitem
\bibitem[Yamaletdinov \latin{et~al.}(2018)Yamaletdinov, Ivakhnenko,
  Sedelnikova, Shevchenko, and Pershin]{Yamaletdinov18}
Yamaletdinov,~R.~D.; Ivakhnenko,~O.~V.; Sedelnikova,~O.~V.; Shevchenko,~S.~N.;
  Pershin,~Y.~V. \emph{Scientific Reports} \textbf{2018}, \emph{8}, 3566\relax
\mciteBstWouldAddEndPuncttrue
\mciteSetBstMidEndSepPunct{\mcitedefaultmidpunct}
{\mcitedefaultendpunct}{\mcitedefaultseppunct}\relax
\EndOfBibitem
\bibitem[Khantha \latin{et~al.}(2004)Khantha, Cordero, Molina, Alonso, and
  Girifalco]{Khantha2004}
Khantha,~M.; Cordero,~N.~A.; Molina,~L.~M.; Alonso,~J.~A.; Girifalco,~L.~A.
  \emph{Physical Review B - Condensed Matter and Materials Physics}
  \textbf{2004}, \emph{70}, 1--8\relax
\mciteBstWouldAddEndPuncttrue
\mciteSetBstMidEndSepPunct{\mcitedefaultmidpunct}
{\mcitedefaultendpunct}{\mcitedefaultseppunct}\relax
\EndOfBibitem
\bibitem[Tsai \latin{et~al.}(2015)Tsai, Chung, Lin, and Yamada]{Tsai2015b}
Tsai,~P.-c.; Chung,~S.-C.; Lin,~S.-k.; Yamada,~A. \emph{Journal of Materials
  Chemistry A} \textbf{2015}, \emph{3}, 9763--9768\relax
\mciteBstWouldAddEndPuncttrue
\mciteSetBstMidEndSepPunct{\mcitedefaultmidpunct}
{\mcitedefaultendpunct}{\mcitedefaultseppunct}\relax
\EndOfBibitem
\bibitem[Dienes(1950)]{dienes1950frequency}
Dienes,~G.~J. \emph{Journal of Applied Physics} \textbf{1950}, \emph{21},
  1189--1192\relax
\mciteBstWouldAddEndPuncttrue
\mciteSetBstMidEndSepPunct{\mcitedefaultmidpunct}
{\mcitedefaultendpunct}{\mcitedefaultseppunct}\relax
\EndOfBibitem
\bibitem[Persson \latin{et~al.}(2010)Persson, Sethuraman, Hardwick, Hinuma,
  Meng, van~der Ven, Srinivasan, Kostecki, and Ceder]{Persson2010}
Persson,~K.; Sethuraman,~V.~A.; Hardwick,~L.~J.; Hinuma,~Y.; Meng,~Y.~S.;
  van~der Ven,~A.; Srinivasan,~V.; Kostecki,~R.; Ceder,~G. \emph{The Journal of
  Physical Chemistry Letters} \textbf{2010}, \emph{1}, 1176--1180\relax
\mciteBstWouldAddEndPuncttrue
\mciteSetBstMidEndSepPunct{\mcitedefaultmidpunct}
{\mcitedefaultendpunct}{\mcitedefaultseppunct}\relax
\EndOfBibitem
\bibitem[K{\"{u}}hne \latin{et~al.}(2017)K{\"{u}}hne, Paolucci, Popovic,
  Ostrovsky, Maier, and Smet]{Kuhne2017}
K{\"{u}}hne,~M.; Paolucci,~F.; Popovic,~J.; Ostrovsky,~P.~M.; Maier,~J.;
  Smet,~J.~H. \emph{Nature Nanotechnology} \textbf{2017}, \emph{12},
  895--900\relax
\mciteBstWouldAddEndPuncttrue
\mciteSetBstMidEndSepPunct{\mcitedefaultmidpunct}
{\mcitedefaultendpunct}{\mcitedefaultseppunct}\relax
\EndOfBibitem
\bibitem[Ni \latin{et~al.}(2010)Ni, Bu, Zou, Yi, Bi, and Chen]{Ni2010}
Ni,~Z.; Bu,~H.; Zou,~M.; Yi,~H.; Bi,~K.; Chen,~Y. \emph{Physica B: Condensed
  Matter} \textbf{2010}, \emph{405}, 1301--1306\relax
\mciteBstWouldAddEndPuncttrue
\mciteSetBstMidEndSepPunct{\mcitedefaultmidpunct}
{\mcitedefaultendpunct}{\mcitedefaultseppunct}\relax
\EndOfBibitem
\bibitem[Kaneko and Saito(2017)Kaneko, and Saito]{Kaneko2017}
Kaneko,~T.; Saito,~R. \emph{Surface Science} \textbf{2017}, \emph{665},
  1--9\relax
\mciteBstWouldAddEndPuncttrue
\mciteSetBstMidEndSepPunct{\mcitedefaultmidpunct}
{\mcitedefaultendpunct}{\mcitedefaultseppunct}\relax
\EndOfBibitem
\end{mcitethebibliography}

\newpage

	\newpage
    \setcounter{page}{1}
    \renewcommand{\thefigure}{S.\arabic{figure}}
    \setcounter{figure}{0}
    \renewcommand{\thetable}{S.\arabic{table}}
    \setcounter{table}{0}
	\section*{Supporting Information}
	\begin{figure}[h]
		\includegraphics[width=.32\textwidth]{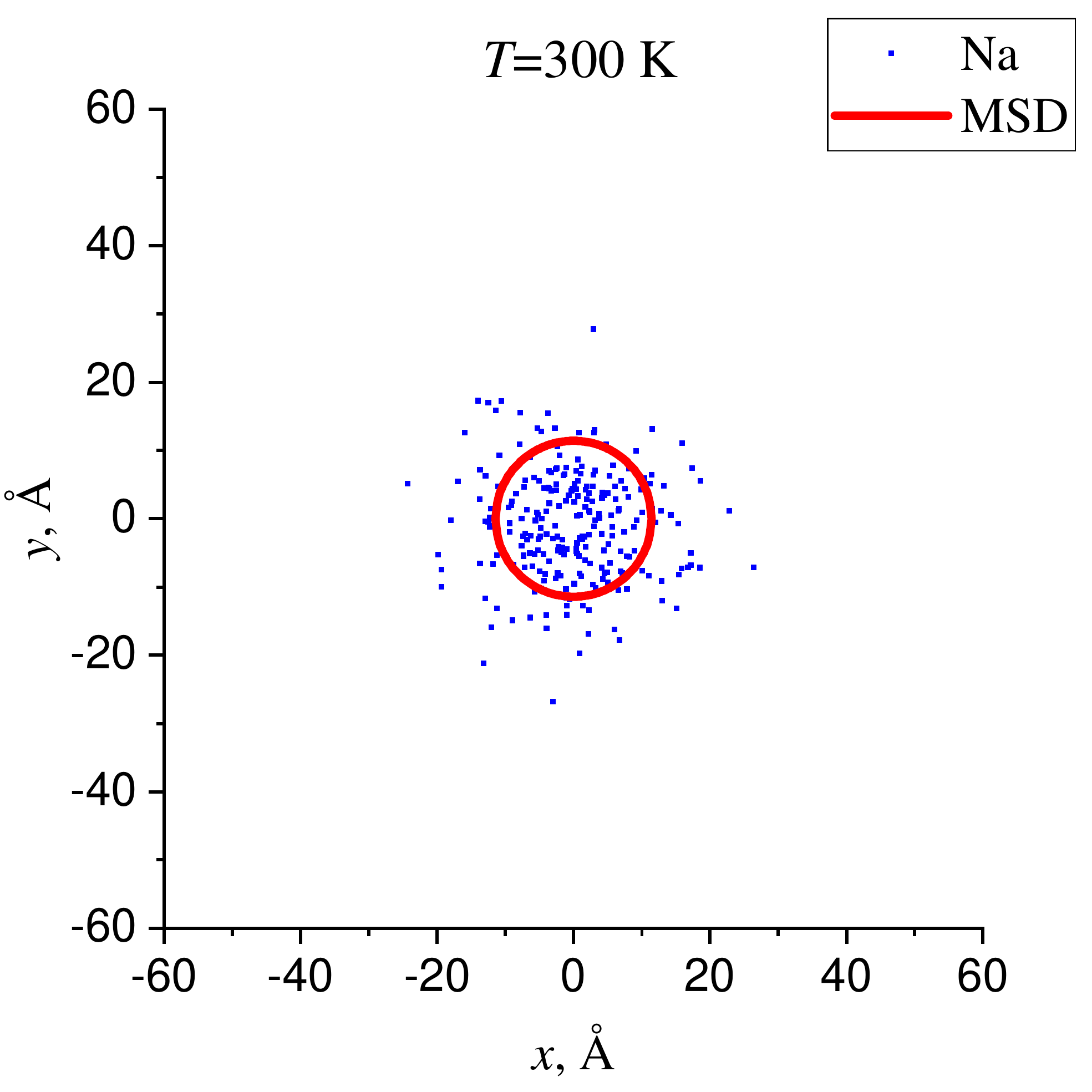}
		\includegraphics[width=.32\textwidth]{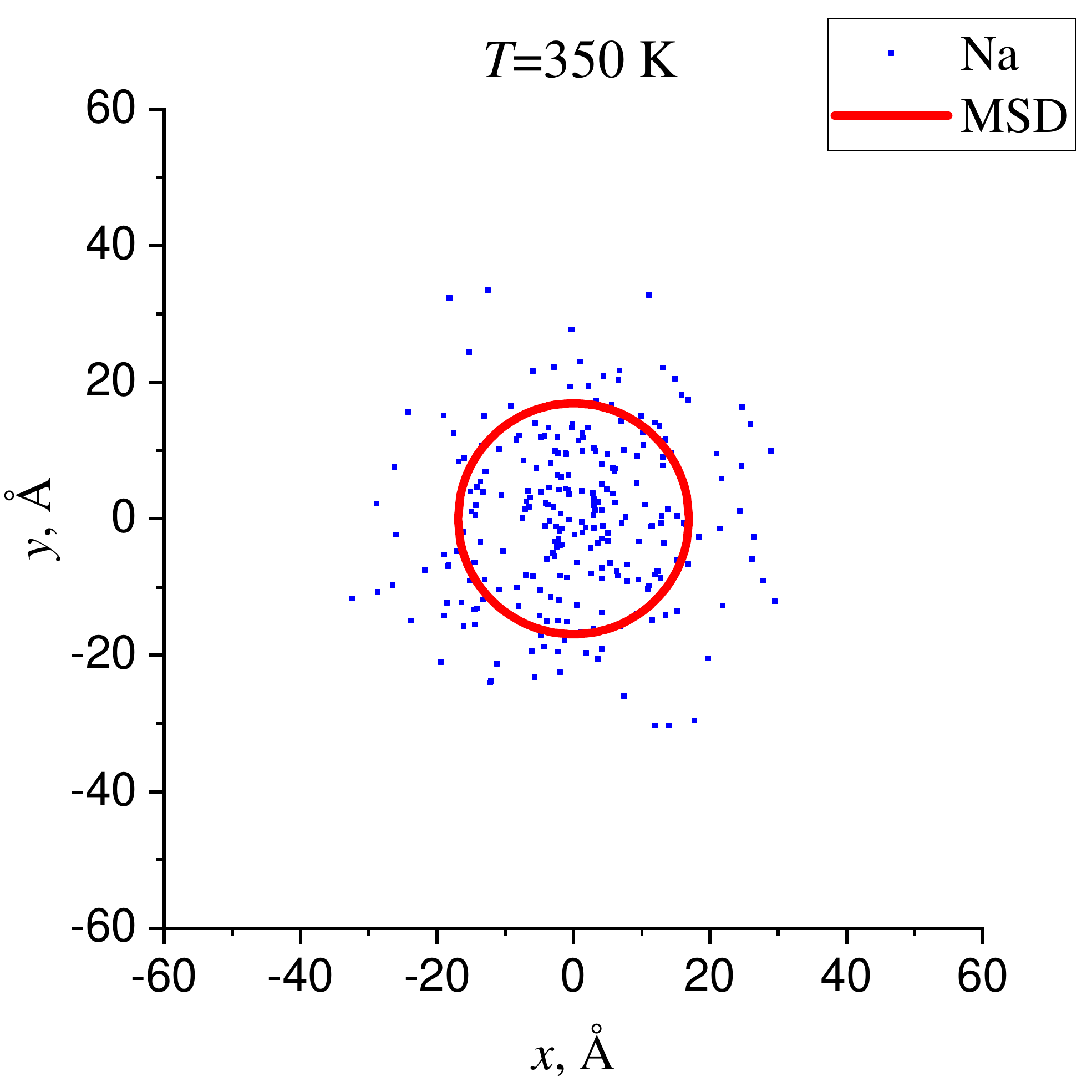}
		\includegraphics[width=.32\textwidth]{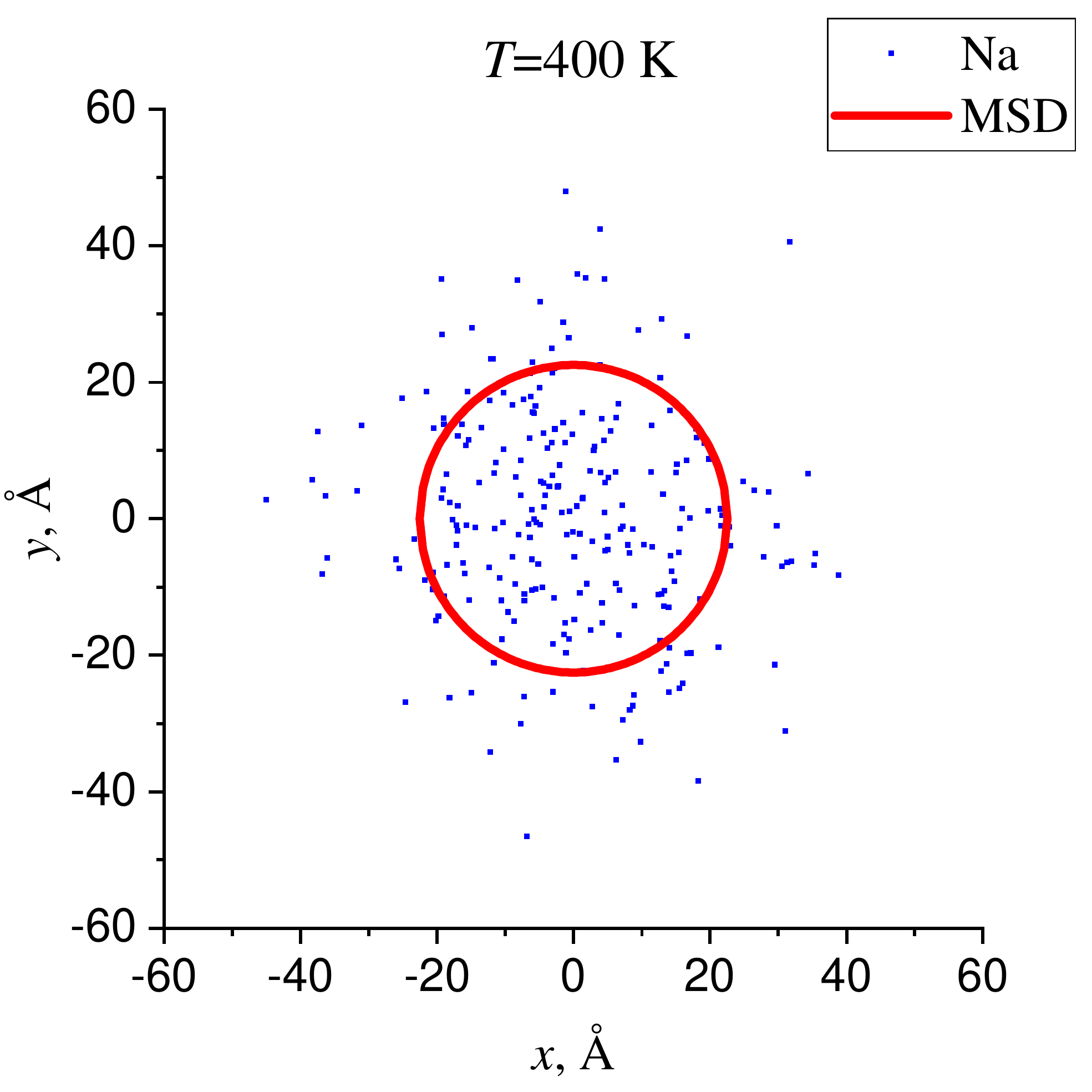}
		\caption{The final displacement of Na atoms in a single-layer graphene at different temperatures. Here, the red circles represent the root MSD of the displacement.}
		\label{fig:S1}
	\end{figure}
	
	\begin{figure}[h]
		\includegraphics[width=.4\textwidth]{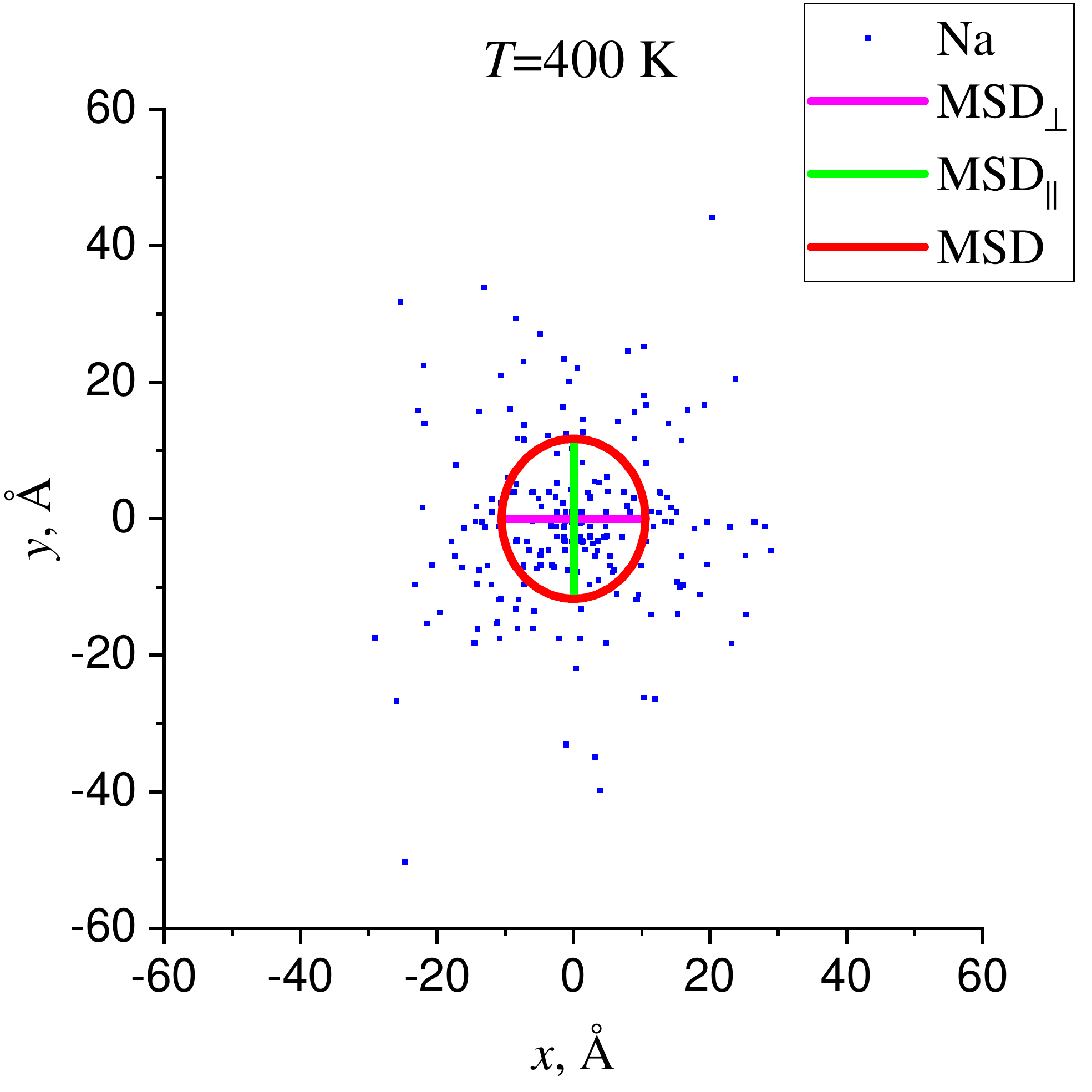}
		\includegraphics[width=.4\textwidth]{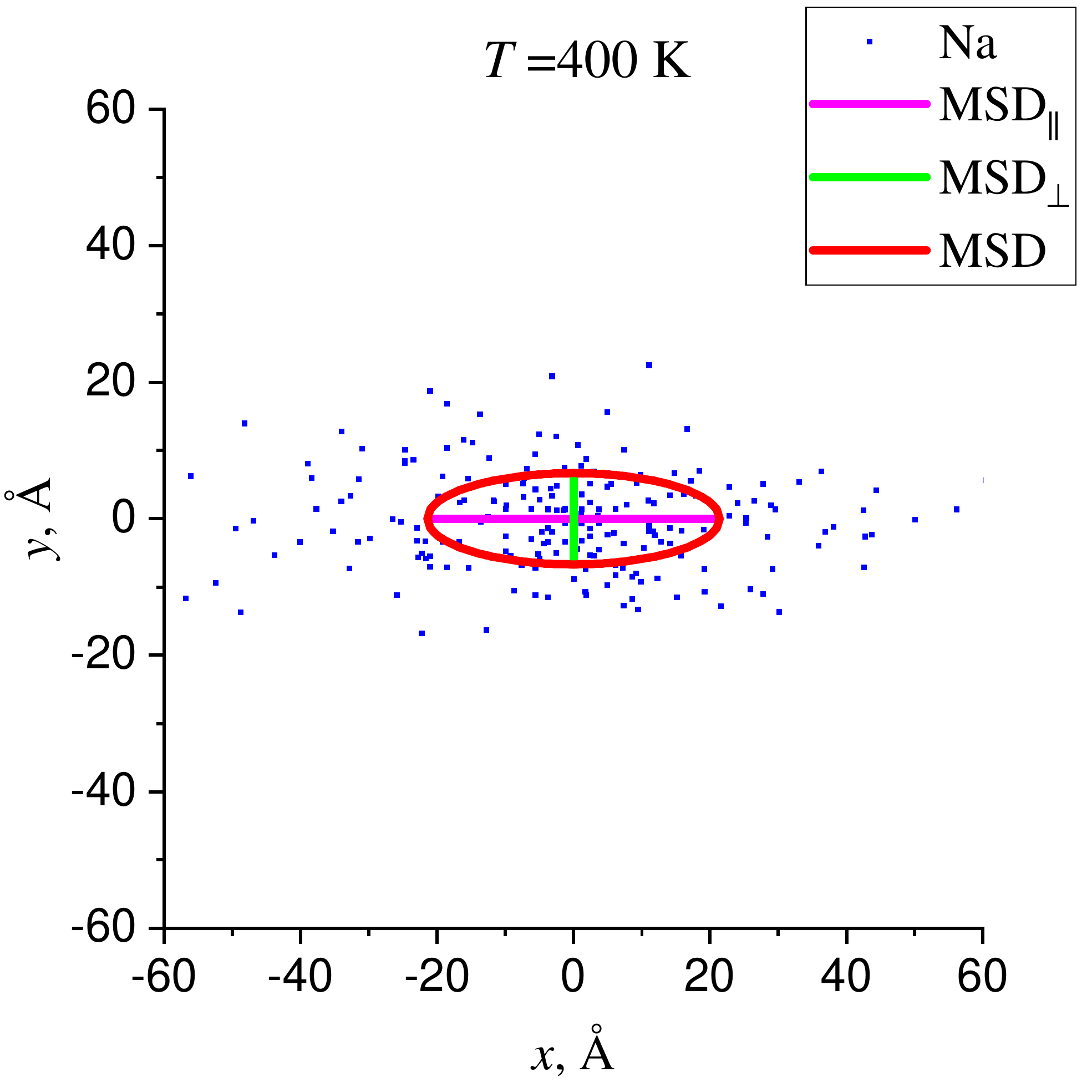}
		\caption{The final displacement of Na atoms in AA-stacked bilayer graphene. The $\parallel$ direction is along the armchair edge (left plot), and zig-zag edge (right plot). The green and purple lines corresponds to the root MSD in $x$ and $y$ directions, respectively.}
		\label{fig:S2}
	\end{figure}

	 \begin{table}[h!]
	 	\begin{center}
	 		\caption{Na diffusion coefficient at $T=300$~K, frequency factor, and activation energy in single-layer (1L) and AA-stacking bilayer (2L) flat graphene.}
	 		\label{tab:diff_isotropNa}
	 		\begin{tabular}{|c||c|c|c||c|}
	 			\hline
	 			system&
	 			$D_{300}$, cm$^2$/s& $Z$, cm$^2$/s&$E$, eV& prior work \\
	 			\hline	 		
	 			Na-1L& $6.5\cdot10^{-6}$& $1.5\cdot10^{-3}$&$0.14$&$E=0.135$ eV~\cite{Koh2013} \\
	 			\hline
	 			AA-Na-2L& $2.5\cdot10^{-5}$& $2.0\cdot10^{-4}$&$0.072$& \\
	 			\hline
	 		\end{tabular}
	 	\end{center}
	 \end{table}

\begin{table}[h!]
		\begin{center}
			\caption{Na diffusion constants at $T=300$~K, frequency factors, and activation energies in AA-stacking bilayer graphene. Here AC (ZZ) denotes the type of the edge that was kept fixed in simulations (the edge in the $\parallel$ direction). The number 90 (or 95) in the system name denotes the degree of buckling (i.e., 95 corresponds to $d/L=0.95$).}
			\label{tab:diff_Na}
			\begin{tabular}{|c||c|c|c||c|c|c|}
				\hline
				system &
				$D_{300}^\parallel$, cm$^2$/s&$Z_\parallel$, cm$^2$/s & $E_\parallel$, eV& $D_{300}^\perp$, cm$^2$/s&$Z_\perp$, cm$^2$/s& $E_\perp$, eV\\
				\hline				
				AA-Na-AC-95& $9.9 \cdot 10^{-6}$ &$0.40\cdot 10^{-4}$ &$0.036$ & $6.8 \cdot 10^{-6}$& $0.56\cdot 10^{-4}$ &$0.055$ \\
				\hline
				AA-Na-AC-90& $9.9 \cdot 10^{-6}$ &$1.7\cdot 10^{-4}$ &$0.074$ & $3.0 \cdot 10^{-6}$& $2.3\cdot 10^{-4}$ &$0.11$\\
				\hline
				AA-Na-ZZ-95& $5.4 \cdot 10^{-6}$& $0.37\cdot 10^{-4}$ &$0.050$& $6.1 \cdot 10^{-6}$ &$0.42\cdot 10^{-4}$ &$0.050$ \\
				\hline
				AA-Na-ZZ-90& $3.1 \cdot 10^{-5}$&$1.2\cdot 10^{-4}$ &$0.034$ &$1.9 \cdot 10^{-6}$ & $0.61\cdot 10^{-4}$ &$0.090$\\
				\hline
			\end{tabular}
		\end{center}
		
	\end{table}
	
\end{document}